\documentstyle[11pt,aaspp4]{article}

\def\etal{{\sl et al.}}
\def\magarc   {{\ mag\ arcsec$^{-2}$} }

\def\mtc      {{ ($m_{AB}^{T}$) } }
\def\lsim{\hbox{ \rlap{\raise 0.425ex\hbox{$<$}}\lower 0.65ex\hbox{$\sim$} }}
\def\gsim{\hbox{ \rlap{\raise 0.425ex\hbox{$>$}}\lower 0.65ex\hbox{$\sim$} }}
\def\cge      {{$_ >\atop{^\sim}$}}

\slugcomment{Accepted to The Astrophysical Journal (Letters)}

\lefthead{Odewahn \etal~  }
\righthead{Host Galaxy of GRB 971214}

\begin{document}

\title{The Host Galaxy of the Gamma--Ray Burst 971214
\footnote{Based on observations with the NASA/ESA {\it Hubble Space Telescope} 
obtained at the Space Telescope Science Institute, which is operated by AURA, Inc., 
under NASA Contract NAS 5-26555; and on observations obtainded at the W. M. Keck 
Observatory which is operated jointly by the California Institute of Technology 
and the University of California.}
}

\author{S. C. Odewahn$^1$, S. G. Djorgovski$^1$, S. R. Kulkarni$^1$}

\author{M. Dickinson$^2$, D. A. Frail$^4$}

\author{A. N. Ramaprakash$^{1,3}$, J. S. Bloom$^1$, K. L. Adelberger$^1$}

\author{J. Halpern$^5$, D. J. Helfand$^5$, J. Bahcall$^6$, R. Goodrich$^7$}

\author{F. Frontera$^8$, M. Feroci$^9$, L. Piro$^9$, and E. Costa$^9$}

\bigskip 

\affil{$^1$ Palomar Observatory 105--24, California Institute of Technology,
            Pasadena, CA 91125, USA}

\affil{$^2$ Allen C. Davis Fellow, Dept. of Physics and Astronomy, The Johns Hopkins 
            University and Space Telescopre Science Institute, 
            Baltimore, MD 21218, USA}

\affil{$^3$ Inter-University Centre for Astronomy and Astrophysics,
            Ganeshkind, Pune 411 007, India}

\affil{$^4$ National Radio Astronomy Observatory, Socorro, NM 87801, USA \footnote{The 
National Radio Astronomy Observatory is a facility of the National Science Foundation
operated under cooperative agreement by Associated Universities, Inc.} }

\affil{$^5$ Dept. of Astronomy, Columbia University, New York, NY 10027, USA} 

\affil{$^6$ Institute for Advanced Study, Princeton, NJ 08540, USA}

\affil{$^7$ W. M. Keck Observatory, 65-1120 Mamalahoa Highway, Kamuela, HI 96743, USA}

\affil{$^8$ Istituto Tec. Studio dell Rad. Extraterrestri, CNR, via Gobetti
            101, Bologna I-40129, Italy}

\affil{$^9$ Istituto di Astrofisica Spaziale, CNR., Via Fermi 21, 
            00044 Frascati, Italy} 

\begin{abstract}

We report on Hubble Space Telescope (HST) observations of the host galaxy of
GRB 971214, taken four months after the burst.  The redshift of the proposed
host galaxy at $z = 3.418$, combined with optical and radio observations of the
burst afterglow, implies the extremely large isotropic energy release from the
burst in $\gamma$-rays of $E_\gamma \approx 3 \times 10^{53}$ erg, some two
orders of magnitude higher than the previously commonly assumed numbers. 
The positional offset between the optical transient observed at the Keck
telescope and the centroid of the proposed host galaxy in the HST image is
$0.14 \pm 0.07$ arcsec.  We find no evidence in our deep HST image for a chance
foreground galaxy superposed along the line of sight to the proposed host at $z
= 3.418$. 
The morphology and photometric properties of this galaxy, such as the
total flux, morphology, radial surface profile and scale length, are typical as
compared to other, spectroscopically confirmed $z\geq3$ galaxies. 

\end{abstract}


\section{Introduction}

The origin of cosmic $\gamma$--ray bursts (GRBs; \cite{kle73}, \cite{fish95}) is 
still one of the outstanding problems in modern astronomy.  
The breakthrough in this field has been the discovery of long-lived x-ray 
afterglows of GRBs (\cite{cos97}, \cite{pir97a},\cite{pir97b}, etc.) using 
the BeppoSAX satellite (\cite{boe97}).  This enabled the discovery of optical 
(e.g., \cite{jvan97}, \cite{bond97}, \cite{djo97}, etc.), and radio 
(e.g., \cite{fra97}, \cite{tay98}, \cite{fra98a}, \cite{fra98b})
afterglows and their subsequent studies.

 Several optical transients (OTs) associated with GRBs have been
found. In the case of GRB 970508 a lower limit to the redshift of 
the OT was obtained, based on intervening absorption systems (\cite{met97}).  
In each case, a faint galaxy was found at the same location (within
a fraction of an arcsecond), after the OT faded.  Redshifts have
been obtained for three such GRB host galaxies: 
$z = 0.835$ for GRB 970508 (\cite{bloom98a}),
$z = 3.418$ for GRB 971214 (\cite{kul98}), and
$z = 0.966$ for GRB 980703 (\cite{djo98d}).
These measurements have established that most or all GRBs are located at 
cosmological distances (\cite{pac95}). 

Perhaps the most spectacular case so far is GRB 971214.  Following the
detection of the burst and its x-ray afterglow 
(\cite{hei97}, \cite{kip97}, \cite{ant97}), an OT was found (\cite{hal98}).
Detailed optical and infrared follow-up observations 
(\cite{kul98}, \cite{ram98}) resulted in the detection of the host galaxy,
measurement of its redshift ($z = 3.418$), and the determination of the
physical parameters of the afterglow.  Little is known concerning 
the geomerty of the detonations producing GRBs and we assume here 
an isotropic release, implying an energy from the burst in 
$\gamma$-rays of $E_\gamma \approx 3 \times 10^{53}$ erg,
some two orders of magnitude higher than the previously commonly assumed
numbers. The high redshift and the spectacular energetics of this burst make it
worthy of a detailed study.  Here we report on Hubble Space Telescope (HST) 
observations of the host galaxy of this burst, about four months after the
burst itself.

\section{Observations and Data Reductions}

HST observations of the GRB 971214 field were obtained on April 13 1998 UT
using the Space Telescope Imaging Spectrograph (STIS, \cite{kim98}). 
The CCD camera was used in imaging mode with the clear (50CCD)
filter.  This provides an essentially unfiltered bandpass defined by the
quantum efficiency of the CCD, covering $\lambda$2000--10000 {\AA } 
with a peak sensitivity at $\lambda$5852 \AA.
A total exposure of 11862 seconds was collected over four consecutive orbits.
Four images were taken per dither position
to facilitate cosmic ray removal.  Initial data processing followed
standard STScI pipeline procedures, including bias and dark current
subtraction, flat field division and cosmic ray removal.  
The dithered images were combined
in two ways to produce science--grade composite images.   We created (i)
a simple median of the image registered on the original STIS detector scale of
$0\farcs05078$/pixel, and (ii) a composite image using the ``drizzling''
algorithm (described, e.g., in \cite{will96} and \cite{fh97}) with a 
scale of $0\farcs033$/pixel.
The background r.m.s.\ per original STIS pixel (gain = 1 elec/ADU) was 
0.0037 ADU/second corresponding to a $1\sigma$ detection threshold per 
pixel of $\sim$26 \magarc.   
Figure 1 shows a portion of the drizzled stack image of the field.  The burst
host galaxy shows a compact, somewhat irregular morphology which is relatively
common for galaxies at comparable magnitudes and redshifts as observed with
the HST.



In order to establish the exact positional relation of the burst afterglow
and the host galaxy, we performed relative astrometry between the drizzled
HST image and ground-based images obtained with the Keck-II 10-m telescope
(\cite{kul98}).  We used the I band
image obtained on 16 December 1997 UT, approximately 1.6 days after the burst,
while the OT was still bright and dominated the total light from the object.
Two different centering algorithms and two different coordinate transformations
were used, with all methods giving consistent results within estimated 
errors.


Comparing the positions of the OT and the host galaxy in the Keck data
alone we find a displacement of $0.11 \pm 0.09$ arcsec. 
Computing the relative positions of the host galaxy as detected in the 
Keck data and in the HST data, we obtain a 
displacement of $0.02 \pm 0.07$ arcsec, i.e., consistent with zero.
For the offset between the OT in the Keck image 
and the host galaxy in the HST image we find a displacement of 
$0.14 \pm 0.07$ arcsec, in an excellent agreement with the Keck 
measurements of \cite{kul98}. The computed position of the OT in 
the HST image is shown in Fig. 2. The offset from the apparent 
nucleus of the host is only a $2\sigma$ result, however the OT position 
lies at a position angle coincident with that of the extended "disk-like" 
extensions of the host, adding weight to the argument that the GRB 
occurred in the outer portion of the host galaxy. 

The surface density of objects in our STIS field that are at least as bright 
as the proposed host is 136 arcmin$^{-2}$. The probability of having a chance 
positional coincidence within 0.2 arcsec is $p = 1.5 \times 10^{-3}$, 
in excellent agreement with the Keck measurements (\cite{kul98}).  
Moreover, even with the superior angular resolution and depth of 
the HST data we see no sign of a possible foreground galaxy superposed 
on the image of the proposed host. We thus conclude that this 
galaxy is indeed the host of GRB 971214. 



Because of the extremely broad bandpass used here, we cannot place our 
measured STIS photometry unambiguously on any standard photometric system.
Given a typical spectral energy distribution
for distant star forming galaxies (\cite{ste96a}) 
and the bandpass of the 50CCD system, the AB magnitudes provided here
probably correspond roughly to those that would be measured through
standard $V$ or $R$ bandpasses. Surface photometry of the host galaxy was 
performed on the median and drizzle images using the MORPHO package (\cite{obw97}).  
Radial surface brightness profiles computed from both the median and 
drizzle composites are shown in Fig. 3 along with the STIS stellar PSF 
modeled by \cite{rob98} for the 50CCD filter. The host galaxy is clearly 
well resolved, although the question of how best to decompose the light 
profile remains open. In the lower panel of Fig. 3 we plot the profile in 
$r^{1/4}$ coordinates, and a significant downturn at small radii, even relative 
to the PSF curve, suggests that the profile may be dominated by an 
exponential profile. 
Integrating the extrapolated surface brightness profiles gives a total magnitude 
 \mtc of 26.62 and 26.61 for the median and drizzle images respectively.
Correcting for a negligible amount of foreground galactic 
extinction (A$_{B} \approx 0.1$ mag, \cite{bh84}) we adopt a total 
AB magnitude of 26.5. The effective radius of the host galaxy was 
determined to be 0.16$''$ $\pm 0.02$ and was derived from the elliptically 
averaged growth curves from the median and drizzled images. 
The e-folding scale length of the profile was determined to lie in 
the range 0.15$''$ to 0.19$''$.


\section{Discussion}

As already noted by \cite{kul98}, the spectroscopic and photometric properties 
of the host galaxy, i.e., its luminosity, star formation rate, etc., are
typical for the field galaxies at comparable redshifts 
(\cite{ste96a}, \cite{ste96b}, \cite{ste98}).  The HST data enable us to
make additional comparisons in terms of the morphology.

We will assume a standard Friedman model cosmology with $H_0 = 65$ km s$^{-1}$
Mpc$^{-1}$, $\Omega_0 = 0.3$, and $\Lambda_0 = 0$.  For $z = 3.418$, the
luminosity distance is $9.7 \times 10^{28}$ cm, and 1 arcsec corresponds to
7.83 physical kpc or 34.6 comoving kpc in projection.  
In this cosmology, the observed luminosity of the galaxy is $\sim L_*$,
in terms of the extrapolated continuum in the restframe B band as compared
to the present-day B band luminosity function (\cite{kul98}).  
At $z=3.4$ our STIS bandpass is measuring rest-frame UV wavelengths, and hence is 
dominated by light contributed by young massive stars. Using a K-correction 
to estimate the corresponding rest-frame optical luminosity is uncertain due 
to our present lack of data on the SEDs and dust absorption properties of $z$\cge3 
galaxies. Following \cite{dic98} we choose to quote the luminosity 
in the UV rest-frame. 
In terms of the far-UV luminosity function as observed among the spectroscopically
confirmed galaxies at $z \sim 3$, the host has an observed luminosity  
$\sim 0.2L_*$ (\cite{dic98}). The star formation rate not corrected for 
any absorption is $\sim 5 ~M_\odot$ yr$^{-1}$ (\cite{kul98}).
Accounting for the typical extinction measured in galaxies at comparable
redshifts (\cite{dic98}) would probably increase these numbers by a factor 
of a few, although the relatively strong L$\alpha$ line emission argues 
against a very heavy extinction. 

Our measured effective radius for the light distribution in the host galaxy
is $\approx 1.3$ kpc.  This is typical 
for the spectroscopically confirmed normal galaxies at comparable redshifts
(\cite{giav96}, \cite{pas96}), and is also comparable to the effective radii
of moderate-luminosity bulges at low redshifts.  Moreover, the morphology of a
compact core surrounded with a slightly irregular envelope, as observed in
this case, is also commonly seen among the field galaxies at $z > 3$
(\cite{giav96}).
On the whole, the physical properties of the host galaxy of GRB 971214 appear
to be typical of the field galaxy population at comparable redshifts.
These objects are now generally interpreted as the progenitors of normal,
$L \sim L_*$ galaxies today, or at least as young bulges of normal spirals.
The observed star formation rate in this object is also typical for normal
galaxies at $z \sim 3$, and is not excessively high.  While it is plausible
that the physical origin of GRBs is somehow related to massive star formation,
we see no evidence that GRBs favor extreme starbursts.


 The small, but significant positional offset between the OT and the host
galaxy nucleus, also observed in the case of GRB 970228 (\cite{sahu97}),
argues against models where GRBs originate from active galactic nuclei or
massive central black holes.  On the other hand, the offsets are sufficiently
small so that the GRB progenitors are likely to be gravitationally bound to
their host galaxies and may not be old systems with large spatial velocities
(see \cite{bloom98b}). Clearly, more OT-host offset measurements made with 
high resolution HST images will be needed to convincingly address this issue.   

 There are two apparently foreground galaxies within about $5''$ of the OT,
one of them at $z = 0.502$ (\cite{kul98}). It is in principle possible 
that the OT was ejected from one of these galaxies and superposed to 
within $0.2''$ on a background galaxy at $z = 3.418$, although as we've 
already pointed out this is statistically very unlikely.  However, there is 
clear evidence for extinction in the broadband spectrum of the OT (\cite{ram98},
\cite{hal98}). This would be hard to understand if the OT were located in 
the outer halo of a $z \sim 0.5$ galaxy, but it is easily understood if 
it were located within the inner portion of an actively star-forming galaxy. 
The hosts for all three of the
GRBs 970508, 971214, and 980703 form a pattern in which the optical transient
is found within the boundaries of a moderate to large redshift host.  Hence,
it is unlikely that one of the galaxies with a separation $\sim 5''$ from 
the GRB 971214 OT could be the host. 


  Finally, we note that a host galaxy was found in every case when either an 
OT was detected or a fading radio transient was observed.  
Such host galaxies of GRBs observed to date have
comparably faint magnitudes, $R \sim 25^m \pm 1^m$ 
(\cite{sahu97}, \cite{bloom98a}, \cite{djo98a}, \cite{djo98b}, \cite{djo98c}).
This removes the so-called ``no host galaxy problem'' for GRBs.  Moreover,
given that most or all of them are likely to be at high redshifts, $z \sim 1$
or greater, there is no need to invoke the hypothesis that GRBs somehow
favor dwarf galaxies.

\acknowledgments

We are grateful to R. Williams of STScI for the allocation of the Director's
Discretionary time for this project, and to the entire BeppoSAX team for their
efforts.  This work was supported in part by the HST grant GO-7964, 
the Bressler Foundation (SGD, SCO), and grants from NSF and NASA (SRK).

\clearpage

\begin{figure}
\figurenum{1}

\plotone{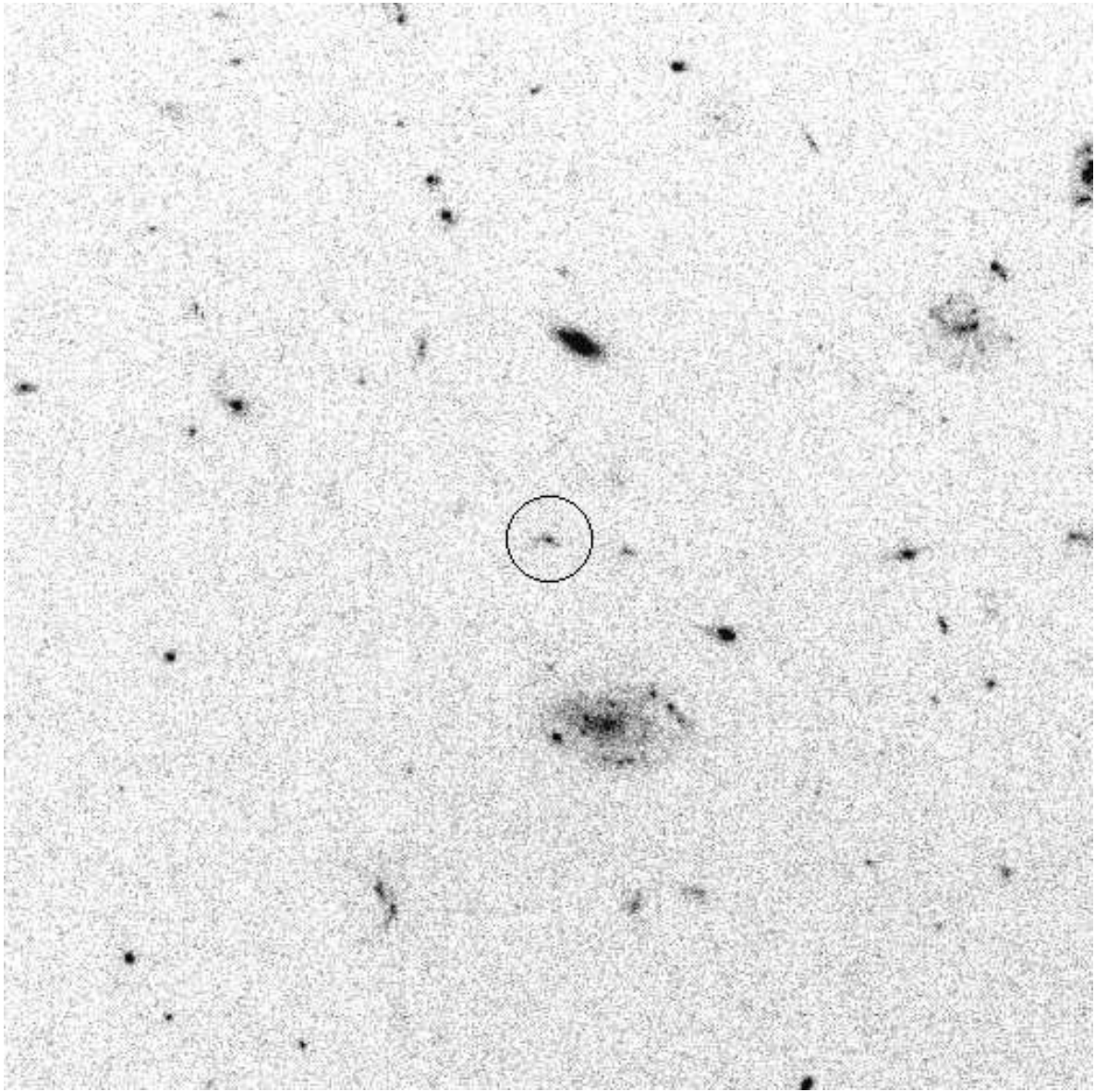}
\epsscale{1.0}

\caption{
A portion of the HST STIS drizzled image of the field of GRB 971214.  
The field shown is 26.0 arcsec square, corresponding to about 200 physical
kpc (900 comoving kpc) in projection at $z = 3.418$ 
(for $H_0 = 65$ km s$^{-1}$ Mpc$^{-1}$ and $\Omega_0 = 0.3$). 
The image is rotated with respect to the standard orientation, so that the
top of the image (vertical axis) is at $PA = 22.5^\circ$.
The burst host galaxy is marked with the circle.
Its coordinates are: $\alpha = 11^h 56^m 26.49^s$,
$\delta = +65^\circ 12^\prime 00.6^{\prime\prime}$ (J2000).
}

\label{fig1}
\end{figure}

\clearpage

\begin{figure}
\figurenum{2}

\plotone{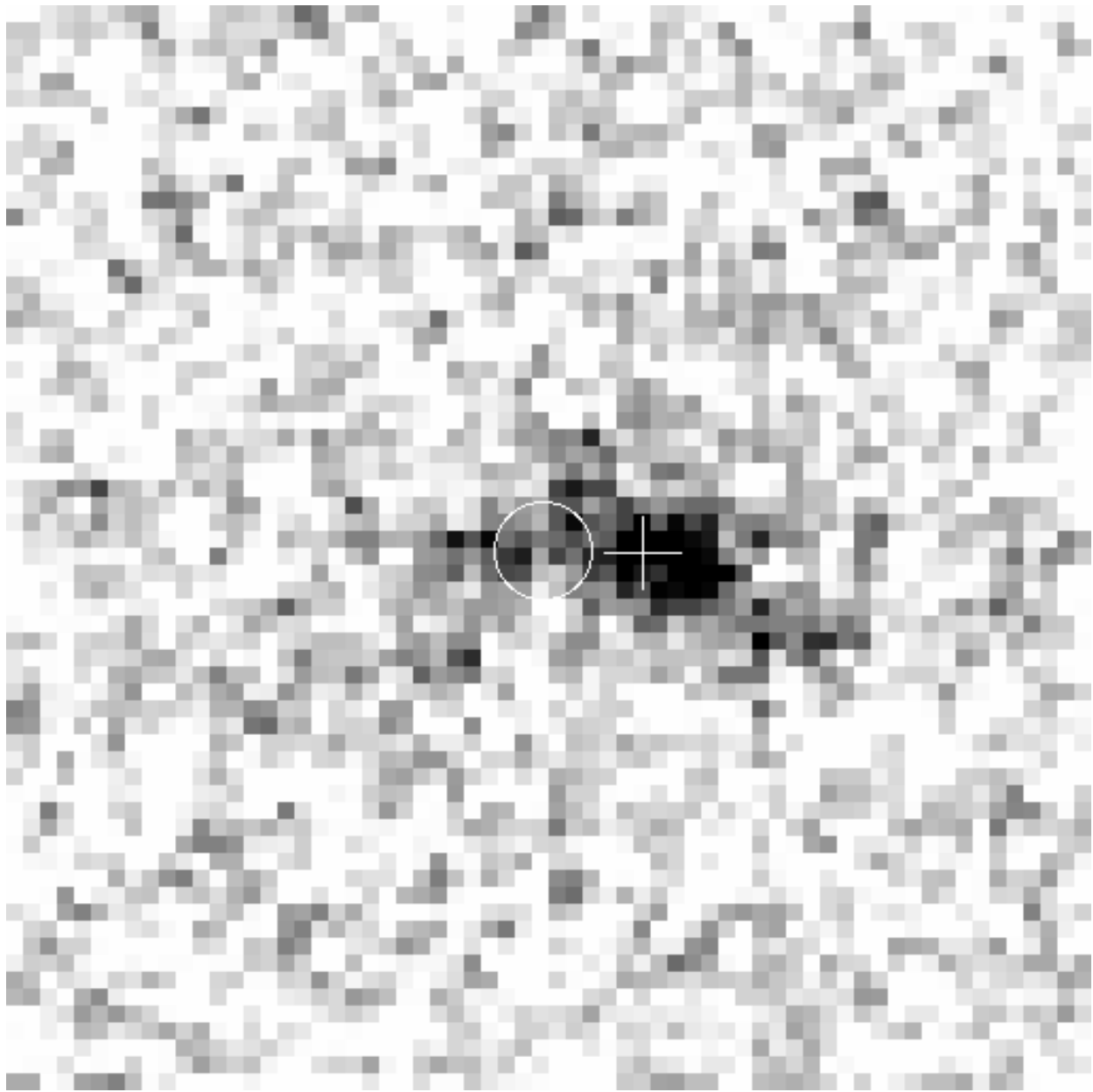}
\epsscale{1.0}

\caption{
A zoom-in on the image shown in Fig.~1, showing the details of the host galaxy.
The field shown is 1.625 arcsec square, corresponding to about 12.7 physical
kpc (56.2 comoving kpc) in projection at $z = 3.418$ 
(for $H_0 = 65$ km s$^{-1}$ Mpc$^{-1}$ and $\Omega_0 = 0.3$). 
The intensity-weighted centroid of the galaxy is marked with the cross, and the
derived position of the OT with the circle; its size corresponds to 1-$\sigma$
errors of our relative astrometry.  Note the irregular morphology of the
galaxy's envelope, and the compact central region (a proto-bulge?). 
}

\label{fig2}
\end{figure}

\clearpage

\begin{figure}
\figurenum{3}

\epsscale{0.5}
\plotone{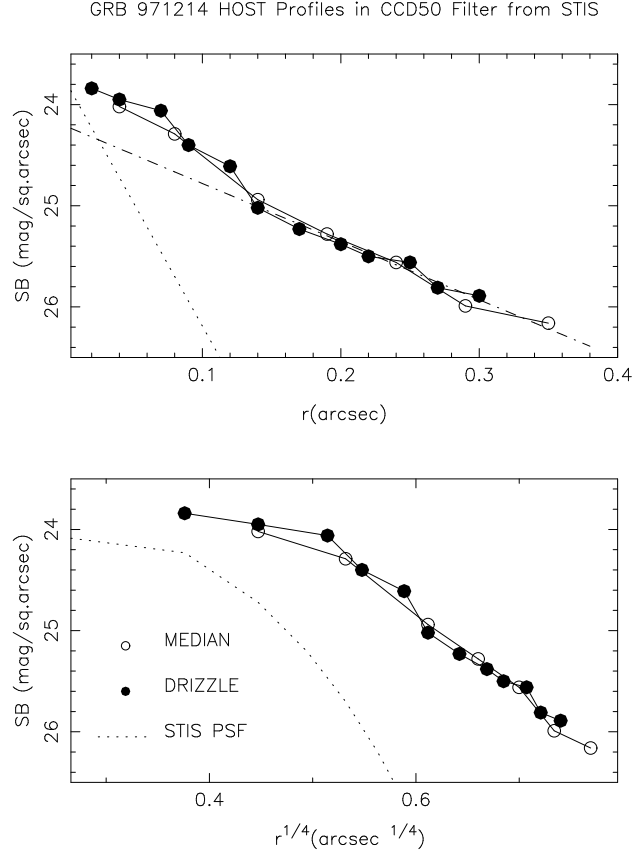}

\caption{
Surface brightness profiles of the GRB host galaxy, derived from the drizzled
image (solid octagons) and from the median stack image (open circles).  The
top panel plots the logarithm of surface brightness (in magnitudes per square
arcsec) against the radius in linear units; an exponential disk profile would
be a straight line on this plot.  The bottom panel shows the same data plotted
against the radius to 1/4 power; a de Vaucouleurs spheroid profile would be
a straight line on this plot.  Both fitting laws are acceptable.  The dotted
lines show the STIS PSF profile modeled by \cite{rob98}, normalized to a 
central SB of 23.74 \magarc to match the host profile central brightness. 
The long-dashed line represents our fit of an exponential for $r\geq0.12$ 
giving an e-folding scale length of 0.19$''$ or 1.3 kpc for our adopted 
cosmology. 
}

\label{fig3}
\end{figure}




\end{document}